\documentstyle[12pt]{article}
\begin{document}
\title {SINGLE-MODE CAVITY-QED OF A RAMAN INTERACTION}
\author { N. Nayak \\
S. N. Bose National Centre for Basic Sciences, \\
Block-JD, Sector-3, Salt Lake City, Calcutta-700091, India \\
S. S. Hassan \\                
Department of Mathematics, Faculty of Science \\
Ain Shams University, Cairo, Egypt \\
and \\
T. Biswas \\
Department of Physics, \\
Indian Institute of Technology, Kanpur-208016, India.\\
}
\newpage
\baselineskip=24pt
\date{}
\maketitle
\begin{abstract}
We consider a single Rydberg atom having two degenerate levels
interacting with the radiation field in a single-mode ideal
cavity. The transition between the levels is carried out by a
$\Lambda$-type degenerate two-photon process via a third level
far away from single-photon resonance. At the start of interaction,
the atom is considered to be in a coherent superposition of its
two levels and the field in a coherent state. We study the dynamics
of the atomic as well as the field states. The squeezing in
the quadratures of atomic states can reach up to $100\%$. The
cavity field evolves to a statistical mixture of two coherent fields
with the phase difference between them decided by the interaction time.
Analysis of entropies of the atom and the field shows that the
two systems are dis-entangled periodically in certain cases.
\end{abstract}
\newpage   
\noindent
{\bf 1. Introduction:}\\
The high-Q cavity quantum electrodynamics (QED) has been extensively
investigated as this simple system can be a source of nonclassical 
fields in addition to answering some fundamental questions in quantum
mechanics. The dynamics involves a single atom with its two or
three Rydberg levels interacting with the cavity field. In the case
of a two-level atomic system, it is the Jaynes-Cummings model (JC) [1]
and the revival in the atomic population in its evolution, 
a singature of quantum mechanics, has been experimentally
investigated [2]. These revivals are, in general, Gaussian in shape
which gets broader in successive appearances in time and finally
overlap with one another giving rise to a chaotic evolution. The
three-level atomic system involving a two-photon process on the
other hand gives rise to compact revivals which are regularly               
placed and are also more distinct for a relatively longer time 
compared to the case in a two-level system [3-6]. A two-photon process
concerns transition from one level to another via an intermediate 
level involving a single photon in each transition. 
If this intermediate level is removed far from    
one-photon resonance, then the three-level system can be reduced to
an effective two-level system, the validity of which has been discussed
in detail in refs. 5 and 6. This process in a high-Q cavity, the      
so-called two-photon micromaser, has been experimentally demonstrated [7]. 
Various nonclassical properties, such as squeezing [8], has also been
theoretically predicted in the two-photon case.
The quadrature squeezing [8a] in a degenerate
two-photon process [9,10] where the involved two photons are of same
frequency can go upto 75 percent. The two-mode squeezing [8b] takes place 
in the case of nondegenerate two-photon process [5,6,11] where the involved
two photons are of different frequencies. Quadrature squeezing [8a] has
also been indicated in the nondegenerate case [12]. In addition to various
squeezing in the radiation field, the atomic states can also be squeezed 
in the two-photon dynamics [13]. In a typical cavity QED experiment,
atom in a selected state enter the cavity at such a rate that at most
one atom is allowed to interact with the field for a fixed duration. 
Hence, the two-photon cavity-QED can be a useful source for obtaining
atoms in squeezed states. In ref. 13, the atom initially in one of the
two degenerate levels connected by a Raman-type interaction with a 
single mode of the cavity field was considered. The study showed that 
squeezing in a particular quadrature of the atomic states can only be possible.
However, this situation can be changed if the atom is in a coherent 
superposition of two levels at the start of the interaction. Atomic squeezing 
can be possible in either quadrature for wider values of involved
parameters and, also, the squeezing can be enhanced. It is also 
interesting to find how the atomic coherence effect the evolution of the  
cavity field. The present paper addresses to these problems and, also,    
examines other aspects of the dynamics such as entropy which is a
measure of atom-field correlation. It is interesting to find if the
atom and the field are dis-entangled during the evolution. In section
2, we present the model with its solution. Sections 3 and 4 examine
atomic and field dynamics respectively. The enropies of the atom and
the field are discussed in section 5.
We conclude the paper in section 6.  \\  \\
\noindent
{\bf 2. The model and its solutions:}\\
We consider two degenerate Rydberg levels $i$ and $f$ of an atom
interacting with the single mode of frequency $\omega$ of an
ideal cavity (Q=$\infty$). The transition between $i$ and $f$ takes
place by a two-photon process via an intermediate level removed
far away from one-photon resonance. This can be represented by
an effective Hamiltonian, in a frame rotating at $\omega a^
{\dagger}a$,
\begin{equation}
H_{eff}=ga^\dagger a(S^{+} + S^{-})
\end{equation}
where $a$($a^{\dagger}$) is the annihilation(creation) operator
for the radiation field. $S^{+}$ and $S^{-}$ are Pauli pseudospin
operators for the atomic levels $\vert i>$ and $\vert f>$. $g$
is the coupling constant for the two-photon interaction. Effective
Hamiltonian of the type in eq. (1) has been widely used in
literature [3,4,13]. The range of
validity of such Hamiltonians has been closely investigated
in refs 5 and 6. It is not the aim in this paper to re-examine
the range of validity of eq. (1). However, we shall restrict
ourselves in a region where eq. (1) is usually a true 
representation of the interaction.\\
The time evolution of the atom-field wave function
$\vert\psi>$ is given by
\begin{equation}
\vert\psi(t)>=\exp(-iH_{eff})\vert\psi(0)>
\end{equation}
where $\vert\psi(0)>=\vert\psi(0)>_{atom}\otimes\vert\psi(0)>_{field}$
is the initial condition. For the atom, we assume
\begin{equation}
\vert\psi(0)>_{atom}=\cos(Z/2)\vert i>+\exp(iW)\sin(Z/2)\vert f>
\end{equation}
where Z represents the degree of superposition between $\vert i>$
and $\vert f>$ and W is the phase of this superposition. The cavity
field at t=0 is assumed to be in a coherent superposition of photon
number states $\vert n>$ with complex amplitude $\alpha$ given by
\begin{equation}
\vert\psi(0)>_{field}=\exp(-\vert\alpha\vert^2/2)\sum_{n=0}^{\infty}
\frac{\alpha^{n}}{\sqrt{n!}}\vert n>.
\end{equation}
The dressed states of the system, given by $H_{int}\vert\psi_{n}^{\pm}>
=\pm gn\vert\psi_{n}^{\pm}>$ with
$$\vert\psi_{n}^{\pm}>=[\vert i,n>\pm\vert f,n>]/\sqrt{2}$$
have been seen to be convenient for simplifying the eq. (2). A straight
forward calculation gives
\begin{eqnarray}
\vert\psi(t)>&=&\exp(-\vert\alpha\vert^{2}/2)\sum_{n=0}^{\infty}
\frac{\alpha^{n}}{\sqrt{n!}}[(\cos (Z/2)\cos gnt -ie^{iW}
\sin (Z/2)\sin gnt )\vert i,n>
\nonumber \\
& & +(e^{iW}\sin(Z/2)\cos gnt -i\cos(Z/2)\sin gnt )\vert f,n>]
\end{eqnarray}
Equation (5) represents combined atom-field system at any time during
the evolution. From the density operator $\rho_{atom-field}=\vert
\psi(t)><\psi(t)\vert$, we can obtain atomic as well as field statistics
by appropriate tracing.\\ \\
\noindent
{\bf 3. Atomic statistics:}\\
The atomic probabilities $\rho_{i,i}(t)$ and $\rho_{f,f}(t)$ for
the states $\vert i>$ and $\vert f>$ respectively have interesting
properties. We have
\begin{equation}
\rho_{i,i}(t)=\sum_{n=0}^{\infty}\vert<n\vert A_{i}(t)>\vert^{2}
\end{equation}
and
\begin{equation}
\rho_{f,f}(t)=\sum_{n=0}^{\infty}\vert<n\vert A_{f}(t)>\vert^{2}
\end{equation}
where
\begin{eqnarray}
\vert A_{i}(t)>&=&\frac{1}{2}\exp(-\vert\alpha\vert^{2}/2)\sum_{n=0}
^{\infty}\frac{\alpha^{n}}{\sqrt{n!}}[e^{ignt}\{\cos(Z/2)- e^{iW}
\sin(Z/2)\}
\nonumber \\
& & +e^{-ignt}\{\cos(Z/2)+ e^{iW} \sin(Z/2)\}]\vert n>
\end{eqnarray}
and
\begin{eqnarray}
\vert A_{f}(t)>&=&\frac{1}{2}\exp(-\vert\alpha\vert^{2}/2)\sum_{n=0}
^{\infty}\frac{\alpha^{n}}{\sqrt{n!}}[e^{ignt}\{e^{iW}\sin(Z/2)-
\cos(Z/2)\}
\nonumber \\
& &+e^{-ignt}\{\cos(Z/2)+e^{iW}\sin(Z/2)\}]\vert n>.
\end{eqnarray}
Examining $\vert A_{i}>$ and $\vert A_{f}(t)>$, we find that for
$gt=\pi/2$ and $Z=W=0$
\begin{equation}
\vert A_{i}>=[\vert i\alpha>+\vert -i\alpha>]/2
\end{equation}
and
\begin{equation}
\vert A_{f}>=-[\vert i\alpha>-\vert -i\alpha>]/2
\end{equation}
are, apart from a normalizasion factor, that for even and odd coherent
states respectively [14, 15]. $\vert A_{i}>$ and $\vert A_{f}>$
interchange properties for $Z=\pi$. Whereas for $gt=Z=W=\pi/2$, we get
\begin{equation}
\vert A_{i}(t)>=\frac{e^{-i\pi/4}}{2}[\vert i\alpha>+e^{i\pi/2}
\vert -i\alpha>]
\end{equation}
and
\begin{equation}
\vert A_{f}(t)>=-\frac{e^{-i\pi/4}}{2}[\vert i\alpha>-e^{i\pi/2}
\vert -i\alpha>]
\end{equation}
which are that for a Yurke-Stoler coherent state [16]. In general,
we have, for $W=\pi/2$,
\begin{equation}
\vert A_{i}(t)>=\frac{1}{2}e^{-iZ/2}[\vert\alpha e^{igt}>+e^{iZ}
\vert\alpha e^{-igt}>]
\end{equation}
and
\begin{equation}
\vert A_{f}(t)>=-\frac{1}{2}e^{-iZ/2}[\vert\alpha e^{igt}>-e^{iZ}
\vert\alpha e^{-igt}>],
\end{equation}
the phase difference between the superpositions being decided by
the interaction time. Thus we see that the quantities $\vert <n
\vert A_{i}(t)>\vert^2$ and $\vert <n\vert A_{f}>\vert^2$ as function
of n display characteristics of the distribution functions for the
cat-like states for the field. However, the summations in eqs. (6)
and (7) give atomic state probabilities $\rho_{i,i}$ and $\rho_{f,f}$
respectively. On the other hand, such summations are the usual
normalization conditions if the distribution functions were for the
radiation field. This is consistent with the fact that the states
in eqs. (10)-(15) are like various cat-like states apart from a
normalization factor. Similar characteristics have been discussed
in the ref. 4.\\
\indent
These are nonclssical properties involved in the atomis state
probabilities. In addition, the quadratures of the atomic states
show squeezing [8a]. We define
\begin{equation}
S^{x}=[S^{+}+S^{-}]/2
\end{equation}
and
\begin{equation}
S^{y}=[S^{+}-S^{-}]/2i
\end{equation}
which are related by
\begin{equation}
[S^{x},S^{y}]=iS^{z}.
\end{equation}
Hence the variances $(\Delta S^{x})^{2}$ and $(\Delta S^{y})^{2}$
in $S^{x}$ and $S^{y}$ respectively obey the uncertainty relation
\begin{equation}
(\Delta S^{x})^2(\Delta S^{y})^2\geq \frac{1}{4}\vert
<S^{z}>\vert^2
\end{equation}
where $S^{z}=(\vert i><i\vert -\vert f><f\vert )/2$ is the
population difference operator between the levels $\vert i>$ and
$\vert f>$. $(\Delta S^{x})^{2}$ or $(\Delta S^{y})^{2}<\vert
<S^{z}>\vert /2$ indicates squeezing in that quadrature. The percentage
of squeezing is an useful parameter in the estimation of noise
reduction and is given by
\begin{equation}
P=100[1-2(\Delta S^{i})^2/\vert <S^{z}>\vert]\%
\end{equation}
where $i=x$ or $y$.
As the present analysis involves a single-atom dynamics, we have $<S^{+}S^{+}> =
<S^{-}S^{-}>=0$ and hence we have
\begin{equation}
(\Delta S^{x})^{2}=(1-\sin^{2}Z\cos^{2}W)/4
\end{equation}
and
\begin{equation}
(\Delta S^{y})^{2}=[1-\{\sin\xi(t)\cos Z -\cos\xi(t)\sin W\sin Z\}^{2}
\exp(-4\vert\alpha\vert^{2}\sin^{2}gt)]/4
\end{equation}
where
$$\xi=\vert\alpha\vert^{2}\sin(2gt).\eqno(22a)$$
The population difference is given by
\begin{equation}
<S^{z}>=\frac{1}{2}[\cos\xi(t)\cos Z +\sin\xi(t)\sin W\sin Z ]\exp(-
2\vert\alpha\vert^{2}\sin^{2} gt).
\end{equation}
We immediately see that, for $Z=0$ or $=\pi$ indicating that the atom
is in the state $\vert i>$ or $\vert f>$ at $t=o$, there is no
squeezing in the X-component as $(\Delta S^{x})^{2}=1/4$ with
$\vert <S^{z}>\vert\leq 1/2$.
The possibilities of squeezing in the Y-component has been
discussed in the ref. (13). However, the situation can be changed if
the interaction starts with the atom in a superposition of its two
states which indeed produces squeezing in the X-component. This is a
key result in the paper.\\
\indent
Now we look into few interesting cases in which squeezing is possible.
For $W=0$ and $gt=m\pi$ with $m=1,2,3,...$, $<S^{+}>=\frac{1}{2}\sin
Z$ and, hence, there is no squeezing in Y-component. The condition
for squeezing in $S^{x}$ becomes $\cos^{2} Z <\vert\cos Z\vert$ which
is clearly satisfied for $Z\not= n\pi/2$ with $n=0,1,2,.....$. Similar
situation arises for $W=0$ and $gt=(2m+1)\pi/2$, $m=0,1,2,...$ where
the condition for squeezing in the X-component takes the form
$$\cos^{2} Z <\vert\cos Z\vert\exp (-2\vert\alpha\vert^{2})$$
which further reduces to $n\pi - \eta <Z<n\pi +\eta$ where
$\eta =\arccos [\exp(-2\vert\alpha\vert^{2})]$ with $n=0,1,2...$.
There are a few cases in which time evolution of squeezing takes
oscillatory patterns, an example of which is displayed in fig. (1)
around $Z=\pi/2$ and $W=0$. For a fixed $Z=\pi/2$ ans small $W$
such that $e^{iW}=1+iW$, we can write
\begin{equation}
<S^{+}>=[1-iW\cos\xi(t)\exp (-2\vert\alpha\vert^{2}\sin^{2} gt)]/2
\end{equation}
and
\begin{equation}
\vert <S^{z}>\vert =\exp(-2\vert\alpha\vert^{2}\sin^{2} gt)
\vert\sin\xi(t)\vert W/2
\end{equation}
This gives us a condition of squeezing in X-component as
\begin{equation}
gt\not=\frac{1}{2}\arcsin (n\pi /\vert\alpha\vert^{2})
\end{equation}
where $n=0, 1, 2,.....$
and in such a situation the squeezing is nearly 100\% according
to eq. (20). $n=0$ indicates that there is no squeezing in the
initial condition of the atom. Further, from eq. (24), we have
$$(\Delta S^{y})^{2}=(1-W^{2}\cos^{2}\xi(t)\exp
(-4\vert\alpha\vert^{2}sin^{2} gt)]/4\cong 1$$
and hence there is no squeezing in Y-component in this situation.
Apart from these special cases, numerical studies show squeezing
in either quadrature. \\ \\
\noindent
{\bf 3. Field statistics:}\\
The field density operator is given by
\begin{equation}
\rho_{f}=\vert A_{i}(t)><A_{i}(t)\vert +\vert A_{f}(t)><A_{f}(t)\vert
\end{equation}
where $\vert A_{i}(t)>$ and $\vert A_{f}(t)>$ are given by eqs. (8) and
(9). Using eqs. (14) and (15), we find that the coherent field
$\vert\alpha >$ at $t=0$ evolves to
\begin{equation}
\rho_{f}=\frac{1}{2}[\vert\alpha e^{igt}><\alpha e^{igt}\vert +
\vert\alpha e^{-igt}><\alpha e^{-igt}\vert]
\end{equation}
which is a statistical mixture of coherent states $\vert\alpha
e^{igt}>$ and $\vert\alpha e^{-igt}>$ [14]. We see that the phase difference
between the two states is decided by the interaction time. This result
is for $W=\pi/2$ in the initial condition in eq. (3). The field also shows
similar characteristics for other values of $W$. The photon distribution
fuction for the field, obtained from eq. (27)
\begin{equation}
P_{n}=<n\vert\rho_{f}\vert n>=e^{-\vert\alpha\vert ^{2}}\frac
{(\vert\alpha\vert ^{2})^{n}}{n!}
\end{equation}
is that for a coherent field. The Wigner function [17] for $\rho_{f}$,
derived by using the method in refs. [15,18], have the form
\begin{eqnarray}
P_{w}(x,y)&=&\frac{2}{\pi}[(1+\cos W\sin Z)
\exp\{-2(x+k_{1}x_{1}-k_{2}y_{1})^{2}
\nonumber \\
& &  -2(y+k_{1}y_{1}+k_{2}x_{1})^{2}\}+(1-\cos W\sin Z)
\nonumber \\
& & \exp\{-2(x+k_{1}x_{1}+k_{2}y_{1})^{2}
-2(y+k_{1}y_{1}-k_{2}x_{1})^{2}\}]
\end{eqnarray}
where $k_{1}=\cos gt$, $k_{2}=\sin gt$ and $\alpha=x_{1}+iy_{1}$
is the complex amplitude of the coherent field at $t=0$. We see that
the function $P_{w}(x,y)$ is always positive and is twin-peaked in the
complex space with each peak being Gaussian in shape. This is
consistent with the fact that the radiation field, in general,
a statical mixture of two coherent fields [14]. It has been
seen that a field with these characteristics does not
possess squeezing properties in its quadratures which is also the
case in the present investigation. Other properties of such fields
have been analysed in ref. (14).\\   \\
\noindent
{\bf 5. Entropy:}  \\
Study of entropy (for atom/field) as a system dynamical parameter
and as a measure of field-atom correlation has been given for the
JC model [19,20] and some of its generalized forms [21-23]. The
Boltzmann-Gibbs entropy is defined by (scaled by Boltzmann's constant)
\begin{equation}
S=-Tr(\rho \ln \rho )
\end{equation}
where $\rho$ is the system density operator. The entropy $S_{f}$ for the
field, represented by $\rho_{f}$ in eq. (27), is given by
\begin{equation}
S_{f}=-(\pi_{1} \ln \pi_{1} +\pi_{2} \ln \pi_{2})
\end{equation}
where $\pi_{1,2}$ are the eigenvalues of $\rho_{f}$ [20],
\begin{equation}
\pi_{1,2}=\lambda_{11} \pm \exp^{\mp\delta}\vert\lambda_{12}\vert
\end{equation}
ans also
$$\pi_{1,2}=\lambda_{22} \pm \exp^{\pm\delta}\vert\lambda_{12}\vert ,
\eqno(33a)$$
with
$$\lambda_{11}=<A_{i}\vert A_{i}>=1/2 +<S^{z}> ,$$
$$\lambda_{22}=<A_{f}\vert A_{f}>=1/2 -<S^{z}> ,$$
$$\sinh\delta = <S^{z}>/ \vert\lambda_{12}\vert ,$$
$$\vert\lambda_{12}\vert=\vert<A_{i}\vert A_{f}>\vert=\sqrt{R^2+I^2} /2 ,$$
$$R=\sin Z\cos W, $$
$$I=\exp (-2\vert\alpha\vert ^2 \sin^2 gt)[\sin Z\sin W\cos\xi (t)
-\cos Z\sin\xi(t)]$$
and $\xi(t)$ and $<S^{z}>$ are given by eqs.(22a) and (23). Similarly,
for the present two-level atomic structure, its entropy $S_{a}$ is
given by [19],
\begin{equation}
S_{a}=-(\alpha _{1}\ln \alpha_{1} +\alpha_{2}\ln \alpha_{2})
\end{equation}
where
$$\alpha_{1,2}={1\pm 2\sqrt{<S^{z}>^2+\vert\lambda_{12}\vert^2}\over2}.
\eqno(34a)$$
It may be easily verified that $S_{a}=S_{f}$ [20] which is due to the
absence of damping processes. Henceforth, we use the symbol $S$ for
either $S_{a}$ or $S_{f}$. From the above analytical expression for
$S$ we note the following:\\
(i) For $gt=n\pi$ and arbitrary W , $e^{\delta}=(1+\cos Z)/\sin Z$
and hence $S=0$. This emplies that the field and the atom are
decorrelated (dis-entangled) periodically.\\
(ii) For strong fields ($\vert\alpha\vert^2\rightarrow\infty$),
$\pi_{1,2}=\alpha_{1,2}=(1\pm R)/2 $
and hence $S$ is independent of time. \\
\indent
The graph for $S(t)$ for $W=Z=0$ and $\vert\alpha\vert ^2=5$ is
presented in fig. (2). For $Z=\pi/4$ and $W=0$ the graph of $S(t)$ is
similar is shape but the horizontal peak value in different. Within a
semiclassical approximation that ignores the field fluctuations the
effective Hamiltonian in eq. (1) reduces to
\begin{equation}
H_{sc}=2g\vert\alpha\vert ^{2}S^{x}
\end{equation}
The equation of motion for the Bloch vector ${\bf S}(t)=(S^{x},
S^{y}, S^{z})$ is of the form
\begin{equation}
{\bf {\dot{S}}}(t)={\bf \Omega}(t)\times{\bf S}(t)
\end{equation}
with ${\bf \Omega}(t)=(2g\vert\alpha\vert ^{2}, 0,
0)={\bf \Omega}(0)$ being constant in time (note that eqs.(35,
36) are special case of the corresponding equations in ref. [23] for
the resonant two-photon JC model when the Stark shift is neglected).
The time evolution of the Bloch vector with the atom initially in an
atomic coherent state, eq. (3), shows that:\\
(i) For the cases $(W=0, \pi /2, \pi; Z=0), (W=Z=\pi /2)$ and $(W=\pi
/2, Z=\pi /4)$ the vector ${\bf S}(0)$ is orthogonal to ${\bf
\Omega}(0)$
which means that the amplitude of the Bloch vector ${\bf S}(t)$ is
maximum and hence the entropy $S$ is maximum. The numerical results show
that $S(t)$ has maximum value $S_{max}=0.658$ in its periodic
evolution.\\
(ii) For the cases $W=0 (\pi)$ and $Z=\pi /2$, the vector ${\bf S}(0)$ is
parallel(anti-parallel) to the vector ${\bf \Omega}(0)$ which means that
the amplitude of the Bloch vector ${\bf S}(t)$ goes to zero and hence
$S(t)$ has a reduced maximum value. In fact, one can show analytically
from the eq. (33) that for $W=0, \pi$ and $Z=\pi/2$, $\pi_{1,2}=1,0$
and hence $S(t)=0$. \\
(iii) For the cases $W=0,\pi$ and $Z=\pi/4$, $S(0)$ is neither
perpendicular nor parallel to $\Omega(0)$ which indicates
a decrease in $S_{max}$. Numerical results show that $S_{max}=0.41$
which is less that its value in the case (i).\\
\noindent
{\bf 6. Conclusion:}  \\
We have analysed a degenerate Raman process involving two degenerate
Rydberg energy levels of an atom interacting with the radiation field
in the single mode of a cavity with $Q=\infty$. The initial condition for
the cavity field is assumed to be coherent. The atomic statistics
display a rich variety of nonclassical properties if the atom is in a
coherent superposition of the two levels at the start of the
interaction. The squeezing is seen to be possible in either
quadrature for a wide range of
numerical values of the parameters involved. Interesting special
cases are, if the atom is in a state
\begin{equation}
\vert\psi (0)>_{atom}=\frac{1}{\sqrt{2}}[\vert i>+(1+iW)\vert f>]
\end{equation}
with no squeezing in its quadratures initially, the X-quadrature gets
nearly 100\% squeezed during the evolution except at singular
atom-field interaction time given by the condition in eq. (26). On the
other hand, if the atom is in one of the two states at $t=0$, then
squeezing appears in the Y-quadrature. Regarding the field statistics,
we notice that the initial coherent state $\vert\alpha >$ evolves
to a statistical mixture of two coherent states.\\
\indent
Entropy evolution was also examined for various initial conditions and
also discussed within the semiclassical Bloch equations.  The equality
of field and atomic entropies is due to the absence of any dissipative
processes.
For short interaction times,
the ideal cavity approximation $(Q=\infty)$ has been seen to be a good
approximation [24]. But, for arbitrary
time, the following master equation needs to be solved,
\begin{eqnarray}
\dot{\rho }=-i[H_{eff},\rho]-\kappa (1+\bar{n}_{th})(a^{\dagger}a\rho
-2a^{\dagger}\rho a+\rho a^{\dagger}a)
\nonumber \\
-\kappa\bar{n}_{th}(aa^\dagger\rho -2a\rho a^{\dagger}+\rho
 aa^{\dagger})
 \end{eqnarray}
 where $H_{eff}$ is given by eq. (1), $\kappa=\omega/2Q$ is the
 cavity dissipation
 constant and $\bar{n}_{th}$ is the average black-body photons in the
 cavity. It is expected that the field entropy will evolve
 independently towards its maximum and/or steady-state value and the
 atomic entropy will be effected by the cavity dissipation processes
 (cf.[25]. Results of these investigations for the present model and
 other damped cavity-QED systems will be presented later.

\newpage
{\bf Figure Captions:}  \\
Figure 1: Percentage of squeezing in $S^{x}$ for $W=0$ and for
$Z=1.50$ (full), $=1.53$ (broken) and $=1.56$ (dotted).\\
Figure 2: Entropy for the field or atom for $Z=W=0$ and
$\vert\alpha\vert ^{2}=5.0$.
\end{document}